%Paper: gr-qc/9410020
%From: BILGE%TRMBEAM.BITNET@vm.gmd.de
%Date: 17-OCT-1994 15:21:53.41 TST

%\magnification=1200
i\nopagenumbers
\def\t{\textstyle}
\parskip=10pt
%\hsize=14truecm
\null
%\vsize=18truecm
\vglue .7truein
\centerline{\bf PARTICLE CREATION IF A COSMIC STRING SNAPS}
\vskip .7truein
\centerline {A. H. Bilge$^*$}
\vskip .2cm
\centerline {Department of Mathematics,}
\centerline{TUBITAK Marmara Research Center, Gebze Kocaeli, Turkey}
\vskip .5cm
\centerline{ M. Horta\c csu }
\vskip .2cm
\centerline{Physics Department,}
\centerline{TUBITAK Marmara Research Center, Gebze Kocaeli, Turkey}
\centerline{I.T.U., 80626, Maslak,Istanbul,Turkey}
\vskip .5cm
\centerline{ N. \"Ozdemir}
\vskip .2cm
\centerline{Physics Department,}
\centerline{I.T.U., 80626, Maslak,Istanbul,Turkey}
\vskip 4cm

\centerline {\bf Abstract}
  We calculate the Bogolubov coefficients
for a metric which describes the snapping of a cosmic
string.
If we insist on a matching condition for all times {\it and} a particle
interpretation,  we find no particle creation.

\vfill
\noindent
$^*$Present Address: Department of Mathematics, Anadolu
University, Eski\c sehir, Turkey

\vfill\eject
\baselineskip=18pt
\footline={\centerline{\folio}}
\pageno=1

\vskip 1cm
\noindent
{\bf 1. Introduction.}
\vskip .2cm

The possibility that cosmic strings may be the root  of the
mechanism explaining galaxy formation is not still ruled out [1].  We
need further experimental data on the anisotropies in the cosmic microwave
radiation, lensing of quasar images, or gravitational radiation
stemming from the decay of strings to accept or reject this
alternative to inflationary quantum fluctuations [2]. If cosmic
strings exist, they may give rise to vacuum fluctuations which in
turn,  may result in particle production [3-8].

In the presence of a time-like Killing vector, one can define in and
out
states  and calculate the Bogolubov coefficients to see whether particle
production actually occurs.
Another method would be an
approximate
field theory calculation [9] which has been applied to
gravitational
particle production during string formation in [10] with a positive
result.

In this paper we investigate the particle
production during the snapping of a cosmic string and contrast our
result with that of [10]. The background metric is chosen as the
Gleiser-Pullin solution [11], which
describes the snapping of a cosmic string whose ends expand with the
velocity of light.

In Section 2, we set up our notation and discuss the
Klein-Gordon equation for a $C^{(0)}$ metric.
In Section 3,  we solve the Klein-Gordon equation (for a scalar field)
and investigate
whether particle production occurs via the Bogolubov
coefficients.

In this analysis we have  used a classical result which allows the
definition of in and out states in Kasner spaces [12].
The attempts for the computation of the Bogolubov coefficients
are given in Sections 3.1 and 3.2.

 In
Sections 3.1, we obtain a solution in terms of the Hankel functions that
has a particle interpretation, and the
behavior as $t\to \infty$ can be interpreted as particle production, but this
solution is not bounded near the $v=0$ hypersurface.
In Section 3.2, we obtain a continuous solution in terms of hypergeometric
functions
but there is no particle production.

We recall that the Bogolubov coefficient method has been used
in [4] to compute the particle production rates in the case of
the instantanous disappearence (or appearence) of a cosmic string
along the $z$-axis.
In [4], the
author uses a smoothed form a the $C^{(0)}$ metric and  finds an exact solution
for the wave equation.
The in and out states correspond respectively to
a spacetime with a cosmic string and to an empty spacetime. The matching
condition for the solutions of the smoothed wave equation gives nonzero
transmission and reflection coefficients, and this situation is interpreted as
particle production. We note however that in [4] the number of particles
generated and the total energy (obtained by numerical integration) are
proportional to $(\alpha^3(1-\beta^2)^2)$ and $(\alpha^4 (1-\beta^2)^2)$ where
$\alpha$ is the the string formation rate. Hence they give meaningful physical
results only if the increase in the particle production rate is compansated by
a decrease in the resulting conical defect. In the case of a $C^{(0)}$ metric,
$\alpha\to\infty$ to start with, hence one should not expect a finite total
particle number and total energy.

In the Gleiser-Pullin metric that we start with, the usual in and out states
correspond
to the interior of a flat expanding sphere and the exterior of this sphere
with a cosmic string along the $z-$axis. Thus
 our in and out states  (see Eq.3.1.4) should be related to Sahni's results [4]
in the limit $t\to \infty$.
 However we cannot  obtain a continuous solution with apropriate
asymptotic behavior. Even in the case we consider a discontinuous solution that
matches at the top two orders as $t\to \infty$, and obtain
``asymptotic Bogolubov coefficients", these  lead to a constant particle
production rate, hence to an  infinite number of particles. We belive that
these unsatisfactory results are related to using a $C^{(0)}$ metric, since we
would have similar problems in Sahni's approach, if we use a $C^{(0)}$ metric,
as discussed in the previous paragraph.

\vskip 1cm
\noindent
{\bf 2. The scalar Klein-Gordon equation.}

\vskip .2cm
We start with the metric given by Gleiser and Pullin
$$ ds^2 =-4du dv+(u-hv)^2d\varphi^2+(u+hv)^2dz^2\eqno(2.1)$$
where
$h=1$ for $v>0$ and $h=\beta^2$ for $v<0$, $\varphi$ is a periodic coordinate,
i.e $\varphi\in [0,2\pi]$ and the points $\varphi=0$ and $\varphi=2\pi$ are
identified,
$z\in(-\infty,+\infty)$, $u>0$, $u<v$ for $u>0$ and $u>-\beta^2 v$
for $u<0$.  Note that $h(v)$ has a jump discontinuity, hence $h(v)v$ is
continuous, thus the metric in (2.1) is $C^{(0)}$. We can equivalently work
with
 a
$C^{(\infty)}$ form of the metric by letting $h(v)$ to be function that rises
fr
om
$\beta^2$ smoothly on a small interval $v\in(-a,a)$. Then the range of $v$ is
defined to be $h(v)v>u$ for $u>0$ and $h(v)v>-u$ for $u<0$.

The parameter
$\beta$ is a number close to but less than $1$, and is
related to deficit angle of a cosmic string. The Ricci
tensor is equal to zero, but one component of the Weyl tensor is
proportional to a Dirac delta function, which indicates the existence of
an impulsive wave.

The Klein-Gordon equation in these coordinates can be obtained as
$$\eqalign{
F_{uv}-&{1\over (u-hv)^2}F_{\varphi\varphi}
               -{1\over (u+hv)^2}F_{zz}\cr
       &        +{1\over 2}h\left[-{1\over u-hv}+{1\over u+hv}\right]F_u
               +{1\over 2}\left[{1\over u-hv}+{1\over
u+hv}\right]F_v=0\cr}\eqno(2.2)$$

We look for solutions that have $e^{i(m \varphi+k z)}$ dependence, with $m$
integer to satisfy the periodicity condition. Hence we let
$$F\to e^{im\varphi}e^{ikz}  F,\quad m\ \ {\rm integer}.$$
Then (2.2) reduces to
$$\eqalign{
F_{uv}+&{m^2\over (u-hv)^2}F
               +{k^2\over (u+hv)^2}F
               +{1\over 2}h\left[-{1\over u-hv}+{1\over u+hv}\right]F_u\cr
&               +{1\over 2}\left[{1\over u-hv}+{1\over
u+hv}\right]F_v=0.\cr}\eqno(2.3)$$

This equation can easily be solved in the domains $v>0$ and $v<0$. The problem
arises in matching these solutions. We would be primarily interested in
continuous solutions with oscillatory behavior as $u\to \infty$ when $v\to 0$.
In the next section it will be seen that this is not possible:
continuous solutions have wrong asymptotic behaviour and the solutions with
correct asymptotic behaviour can be matched only up to two leading orders.

We can then ask whether we can have generalized solutions, i.e.,
the ones that can have jump discontinuities or $\delta$ function
discontinuites. Note that a solution with no $\delta$ function discontinuity is
bounded. We will show below that if $F$ is bounded, then the $u$ dependence of
the jump discontinuity is given  by Eq.(2.4).

Assume that $F$ is bounded on $(-a,a)$ and $F\mid_{v=-a}=\phi^-(u)$,
$F\mid_{v=a}=\phi^+(u)$. Since $h(v)v$ and its derivative are
bounded, integrating (2.3) from $-a$ to $a$ (using integration by parts for the
last term) and letting $a\to 0$, we obtain
$$(\phi^+_u-\phi^-_u)+{1\over u}(\phi^+-\phi^-)=0\eqno(2.4)$$
Thus $(\phi^+-\phi^-)$ is proportional to ${1\over u}$.

If $F$ is allowed to have $\delta$ function (and derivatives of
$\delta$ functions) discontinuities, then it can be seen that any jump
discontinuity can be accomodated. But as the physical meaning of such solutions
is not well defined, we defer the discussion of such solutions.

In Section 3.1, we obtain solutions with correct asymptotic behavior that can
be matched only up to $O({1\over t})$. Thus the discontinuity across $v=0$
hypersurface obeys (2.4)  asymptotically, hence is bounded as $t\to \infty$.

\vskip 1cm
\noindent
{\bf 3. Computation of the Bogolubov coefficients.}
\vskip .2cm
\noindent
{\bf 3.1. Solution using Hankel functions.}
\vskip .2cm
We can solve the equation (2.3) for $v\ne 0$ in the following
coordinate system.
$$\eqalign{t&=u+\beta^2v\cr
           s&=u-\beta^2v\cr}\eqno(3.1.1)$$
This coordinate system results in incoming and outgoing waves for asymptotic
times and it is appropriate for particle interpretation [12].
 Then the
equation reduces to
$$\left\{\left[{1\over t}{\partial\over \partial t }-
{1\over s}{\partial\over \partial s }+
{\partial^2\over \partial t^2 }-
{\partial^2\over \partial s^2 }\right]+{m^2\over \beta^2s^2 }+
{k^2\over \beta^2 t^2 }\right\}\phi=0\eqno(3.1.2)$$
with solutions
$$Z_m(as)Z_{ik}(at)\quad {\rm for}\ \ v>0 \eqno(3.1.3a)$$
$$Z_{m\over\beta}(as)Z_{ik\over\beta}(at)\quad {\rm for}\ \ v<0 \eqno(3.1.3b)$$
where $a$ is a  constant parameter and the $Z$'s are generic functions
obeying Bessel's
equation. We take solutions
$$\eqalign{
F^{{\rm (in)}}&=\tilde{A}H^{(1)}_m(as)H^{(2)}_{ik}(at)\cr
F^{{\rm (out)}}&=\tilde{B}H^{(1)}_{{m\over \beta}}(as)H^{(2)}_{{ik\over
\beta}}(at)
+\tilde{C}H^{(2)}_{{m\over \beta}}(as)H^{(1)}_{{ik\over
\beta}}(at)\cr}\eqno(3.1.4)$$

If we want a continuous solution we need to match the respective solutions at
$t=s$. We will try to match the solutions asymptotically, we will see that we
can do this only in the leading two orders.

We obtain
$$\eqalignno{
\tilde{A}&e^{i{\pi\over 2}(-m+ik)}
\left[1-{1\over 2it}
{\Gamma(m+\t{3\over 2})\over \Gamma(m-\t{1\over 2})}+O(t^{-2})\right]
\left[1+{1\over 2it}
{\Gamma(ik+\t{3\over 2})\over \Gamma(ik-\t{1\over 2})}+O(t^{-2})\right]\cr
=&\tilde{B}e^{i{\pi\over 2\beta}(-m+ik)}
\left[1-{1\over 2it}
{\Gamma(\t{m\over \beta}+\t{3\over 2})\over \Gamma(\t{m\over
\beta}-\t{1\over
2})}+O(t^{-2}) \right] \left[1+{1\over 2it}
{\Gamma(\t{ik\over \beta}+\t{3\over 2})\over \Gamma(\t{ik\over \beta}-\t{1\over
2})}+O(t^{-2})\right]\cr
+&\tilde{C}e^{i{\pi\over 2\beta}(m-ik)}
\left[1+{1\over 2it}
{\Gamma(\t{m\over \beta}+\t{3\over 2})\over \Gamma(\t{m\over \beta}-\t{1\over
2})}+O(t^{-2}) \right] \left[1-{1\over 2it}
{\Gamma(\t{ik\over \beta}+\t{3\over 2})\over \Gamma(\t{ik\over \beta}-\t{1\over
2})}+O(t^{-2})\right]&(3.1.5)\cr}$$

We define
$$A=\tilde{A}e^{i{\pi\over 2}(-m+ik)},\quad
B=\tilde{B}e^{i{\pi\over 2\beta}(-m+ik)},\quad
C=\tilde{C}e^{-i{\pi\over 2\beta}(-m+ik)},\quad\eqno(3.1.6)$$
and
$$\omega(m,ik)=
-{\Gamma(m+\t{3\over 2})\over \Gamma(m-\t{1\over 2})}
+{\Gamma(ik+\t{3\over 2})\over \Gamma(ik-\t{1\over 2})}\eqno(3.1.7)$$
Then the equation reduces to
$$\eqalign{ A\left[1+\t{1\over 2it}\omega(m,ik)+O(t^{-2})\right]=&
   B\left[1+\t{1\over 2it}\omega(\t{m\over \beta},\t{ik\over
\beta})+O(t^{-2})\right]\cr
  &+ C\left[1-\t{1\over 2it}\omega(\t{m\over \beta},\t{ik\over
\beta})+O(t^{-2})\right].}\eqno(3.1.8)$$
   At the first two orders we obtain
  $$A=B+C$$
  $$A\omega(m,ik)=(B-C)\omega(\t{m\over \beta},\t{ik\over \beta})\eqno(3.1.9)$$
  Using also the condition
  $$\vert B\vert^2-\vert C\vert ^2=1\eqno(3.1.10)$$
  we solve $\vert B\vert^2$ and $\vert C\vert^2$ as
  $$
  \vert B\vert^2={\vert \omega+\tilde{\omega}\vert^2\over
           \vert \omega+\tilde{\omega}\vert^2
           -\vert \omega-\tilde{\omega}\vert^2}\quad\quad
  \vert C\vert^2={\vert \omega-\tilde{\omega}\vert^2\over
           \vert \omega+\tilde{\omega}\vert^2
           -\vert \omega-\tilde{\omega}\vert^2}\eqno(3.1.11)$$
where $\omega=\omega(m,ik)$, $\tilde{\omega}=\omega({m\over \beta},{ik\over
\beta})$.

Using the properties of the $\Gamma$ function it can be seen that
$$\vert C\vert^2=\beta^2\big(1-{1\over \beta^2}\big)^2.\eqno(3.1.12)$$
Thus the transmission coefficient is constant, hence
its integral cannot be finite, and we cannot obtain a finite total number of
particles.

As discussed at the end of Section 1, this result is related to using a
$C^{(0)}$ metric.

\vskip .5cm
\noindent
{\bf 3.2. Solution using hypergeometric functions.}
\vskip .2cm
In this section we obtain a solution which is continuous across $v=0$
hypersurface but asymptotically it goes as a power of $t$. Here we use the
following
unorthodox coordinate system:
$$u=p\cosh\theta\eqno(3.2.1)$$
$$\beta^2v=p\sinh\theta\eqno(3.2.2)$$

We note that, since the Fulling et.al. result [12] is shown for systems with
Hankel function
solutions, a positive result in this coordinate system would not mean particle
production. Despite this remark, we will use our negative result as a
plausibility argument, since it is in line with our previous result.

If we take solutions of the type
$$f=e^{im\phi}e^{i\kappa z}g(p,\theta)\eqno(3.2.3)$$
our equation reads
$$(-p^2{\partial^2\over{\partial p^2}}+2\coth 2\theta
{\partial^2\over{\partial p\partial\theta}}-p{\partial\over
{\partial p}}-{\partial^2\over{\partial\theta^2}}-
{2k^2\over{\beta^2}}(\coth 2\theta-1)+
{2m^2\over{\beta^2}}(\coth 2\theta+1))g=0\eqno(3.2.4)$$
where $\kappa=ik$.
An immediate solution is
$$g=p^s (x^2-1)^{{-s\over 4}+{m-k\over 4}}
[{2\over{(1+x)}}]^{{-s\over 2}+{m-k\over 2}}
e^{-\theta(m+k)} \ _2F_1[{k+m\over 2}-{s\over 2},{m-k\over 2}-{s\over
2},-s;{2\over{1+x}}] \eqno(3.2.5)$$
where $x=\coth 2\theta$ ,$ _2F_1(a,b,c,x)$ is the hypergeometric function.
This solution goes as $t$ to a power in the asymptotic region;
hence lacks the particle interpretation given in  [12].
  When $v=0$, we see that
$\theta=0$ and this reduces to $p^s$.

For $v<0$ we write the solution as,
$$g=p^s(x^2-1)^{{k+m\over{4\beta}}-{s\over 4}}
({2\over{1+x}})^{{k+m\over{2\beta}}-{s\over 2}}e^{-({m-k\over\beta})\theta}\-
_2F_1[{k+m\over{2\beta}}-{s\over 2},{k-m\over{2\beta}}-{s\over 2},-s;
{2\over{1+x}}]\eqno(3.2.6)$$
 This solution also has the power behavior in the asymptotic region.
 This solution reduces to $p^s$ when $v=0$. Although we achieved
the matching without problem with reasonable behaviour at the asymptotic
region,
we see that we can't fix the two independent functions needed.
We know that when $v<0$, we get two independent solutions for the
hypergeometric equation
$$u_1=_2F_1(a,b,c;z)\eqno(3.2.7)$$
$$u_2=z^{1-c}\- _2F_1[a-c+1,b-c+1,2-c;z]\eqno(3.2.8)$$

The second solution is either identically zero or singular at $z$ equals
zero depending on the sign of $(1-c)$. We choose the  former case.
Here $c$ equals one is not
allowed. In this case  matching can be achieved
 when the second independent function  is  put to zero.  This
shows particle production can be consistently put to zero.

\vskip 1cm
\noindent
{\bf 4. Conclusion.}
\vskip .2cm

We find that if we insist on a particle interpretation we can not get a
consistent matching    of our    solutions in a specific $C^{(0)}$ metric.
Only if we neglect    terms    that are $O(t^{-2})$ and further, we
can find the   matching     coefficients which give infinite production.
Note also that a possible matching is possible in   coordinate systems
where a particle interpretation  can  not be given.

We see that we have to be careful in applying approximate method to solutions
for metrics with abrupt changes. We agree with Bernard-Duncan [13] who state
that we
have to
be very careful in problems using $C^{(0)}$
metrics.  We see that any of the methods used in the smoothed
case [4], don't seem to work in our case.
Our result also shows that the results obtained from  approximate  field
theory calculations
[9,10] should be taken with a grain of salt.
\bigskip

\noindent
{\bf Ackowlewdgement:}  We thank Prof.Dr. Yavuz Nutku for suggesting
this problem. This
work is partially supported by T\" UBITAK, The Scientific and Technical Council
of
Turkey under TBAG-\c CG/1.
\vfill\eject

\noindent {\bf REFERENCES}
\vskip .2cm
\baselineskip=14pt
\parskip=0pt
\item{1.}A. Vilenkin, Phys. Reports {\bf C121} (1985) 263.
\item{2.}C.Frenk, S.White, M.Davis and G. Estathiou, Astrophys. J.
{\bf 327} (1982) 507.
\item{3.}L.Parker, Phys.Rev.Lett. {\bf 59} (1987) 1365.
\item{4.}V. Sahni, Modern Phys.Lett.A {\bf 3} (1988) 1425.
\item{5.}G.Mandell and W.Hiscock, Phys.Rev. {\bf D40} (1989) 282.
\item{6.}T.M. Helliwell and K.A. Konkowski, Phys.Rev. {\bf D34}
         (1986) 1908.
\item{7.}B.Linet,Phys.Rev. {\bf D33} (1986) 1833;
         Phys.Rev. {\bf D35} (1987) 536.
\item{8.}N.D.Birrell and P.C.W. Davies, {\it Quantum Fields in Curved
         Space} (Cambridge University Press).
\item{9.}J.Frieman, Phys.Rev. {\bf D39} (1989) 389;
\item{ } Ya. B.Zel'dovich and A.A.Starobinsky JETP
Lett. {\bf 26} (1977) 1252.
\item{10.}V.Hussain, J.Pullin, E.Verdaguer, Phys.
Lett., {\bf B232} (1989) 299.
\item{11.}R.Gleiser, J.Pullin, Class. Quantum Gravity,
    {\bf 6} (1989) L.141.
\item{12.}S.A. Fulling,  L. Parker and B.L. Hu, Phys.
Rev., {\bf D12} (1974) 3905.
\item{13.}C.Bernard and A.Duncan, Annals Phys.
(N.Y) {\bf 107} (1977) 201.
\end